\documentstyle[preprint,aps]{revtex}

\newcommand{\beq}{\begin{eqnarray}}
\newcommand{\eeq}{\end{eqnarray}}

\newcommand{\vx}{\mbox{$\bf r$}}

\newcommand{\vA}{\mbox{$\bf A_{\mit b}$}}

\begin{document}
\title{
\bf Landau Level Mixing and Solenoidal Terms in \\
Lowest Landau Level Currents
}
\author{
R. Rajaraman \cite{rr}
}
\address{
School of Physical Sciences,\\
Jawaharlal Nehru University, \\
New Delhi 110067, INDIA
}
\author{
S.\ L.\ Sondhi
}
\address{
Department of Physics, \\
University of Illinois at Urbana-Champaign, \\
Urbana, Illinois 61801, USA
}
\maketitle

\begin{abstract}
We calculate the lowest Landau level (LLL) current by working in the
full Hilbert space of a two dimensional electron system in a magnetic
field and keeping all the non-vanishing terms in the high
field limit. The answer a) is not represented by a simple LLL operator and
b) differs from the current operator, recently derived by
Martinez and Stone in a field theoretic LLL formalism, by solenoidal terms.
Though that is consistent with the inevitable ambiguities of their Noether
construction, we argue that the correct answer cannot arise naturally in the
LLL formalism.
\end{abstract}

\pacs{PACS numbers:
        73.40.Hm %QHE
}

% TEXT

The microscopic theory of the quantum Hall effect (QHE) \cite{ref:qhe} is
constructed around
a large magnetic field (B) expansion which reflects the fact that the QHE is
a strong field phenomenon. At integer QHE filling factors ($\nu$) the strong
field states are
non-degenerate eigenstates of the kinetic energy---filled Landau levels---and
further terms in the expansion, obtained by perturbing in the interaction,
serve largely to renormalize excitation energies.
At the fractional fillings the eigenstates of the kinetic energy are
macroscopically degenerate and the lowest order solution requires degenerate
perturbation theory in the interaction in a given Landau level. This is the
step where the novel physics arises; inclusion of Landau level mixing is, as
in the IQHE, qualitatively unimportant.
It is worth emphasizing that the irrelevance of LL mixing is
really an example of adiabatic continuity, for the ratio $(e^2/\epsilon l)/
\hbar \omega_c$ ($l=\sqrt{\hbar c/e B}$ is the magnetic length,
$\hbar \omega_c = e B/m c$ is the cyclotron frequency calculated with the band
effective mass and $\epsilon$ is the background dielectric constant) that,
naively, controls the mixing is $O(1)$ in the experiments and is {\em not}
negligible \cite{ref:fn1}.

The solution of the degenerate problem for $\nu <1$, and by extension for other
fractions, is greatly facilitated by restricting the Hilbert space to the
lowest Landau level (LLL) and working with projected operators \cite{ref:gj}.
This
procedure is not without its subtleties for restricting the Hilbert space
restricts the intermediate states that occur in evaluating the product of
a string of operators and hence the product of the projections does not, in
general, equal the projection of the product; the simplest instance of this
effect is that the projected particle co-ordinates obey the equal time
commutator $[x,y]=il^2$.
This letter is concerned with another, unphysical, instance
that was first noted by Girvin, MacDonald and Platzman \cite{ref:gmp}.
If $H_L$, $\rho_L(\vx)$ and ${\bf j}_L(\vx)$ are the projected Hamiltonian,
density and current operators then,
\begin{equation}
\nabla \cdot {\bf j}_L(\vx) =0 \; \; {\rm while} \; \; \partial_t \rho_L(\vx)
 = \frac{i}{\hbar} [H_L,\rho_L(\vx)] \neq 0
\end{equation}
thus violating continuity. These authors noted that for smoothly  varying
potentials $V(\vx)$, $\partial_t \rho_L(\vx)$ could be rearranged into the
divergence of the drift current ${\bf j}_D(\vx) =(c/B^2) ({\bf B} \times
\nabla V(\vx) ) \rho_L(\vx)$.
Subsequently, Sondhi and Kivelson \cite{ref:lri} showed that the drift
current could be obtained as a LL mixing contribution that survives at
arbitrarily high fields. (Readers unfamiliar with this issue may find it
helpful to skim the concluding discussion at this point.)

Recently, Martinez and Stone \cite{ref:ms} used a field theory incorporating
the LLL
constraint to construct a LLL Noether current for the potential scattering
problem that (automatically) satisfies continuity.
Subsequently, Rajaraman \cite{ref:rajaraman} derived the same operator as
well as a form valid for
interacting systems by working directly with the projected equations of
motion as in the usual derivation of the Schr\"odinger current. In this
communication we take a different tack: we calculate the LLL current
by working in the {\em full} Hilbert space of the system and keeping all the
non-vanishing terms in the high field limit. We show that our answer, which
is the correct physical current, differs from the the results in
\cite{ref:ms,ref:rajaraman} by solenoidal terms that do not appear to arise
naturally in the projected problem. Furthermore,
the current in an arbitrary state, generically, has a (solenoidal)
contribution that is not calculable in advance of diagonalizing the projected
Hamiltonian and hence it is not possible to specify a useful form for a
current operator in the LLL that reproduces the exact answer. We return to
the implications of our results at the end.

\noindent
{\bf Projection:} We begin by reformulating, following \cite{ref:lri}, the
high field dynamics as a purely LLL problem; this is an ancient technique,
we provide details solely in the interests of clarity \cite{ref:fn2}.
The Hamiltonian is,
\begin{equation}
H= \frac{1}{\lambda} H_K + V
\end{equation}
where the kinetic term,
\begin{equation}
H_{K}=  \frac{1}{2m} \int d^{2}r \: \psi^{\dagger}(\vx)
{ \left[ \frac{\hbar}{i} \nabla - \frac{e}{c} \vA (\vx) \right] }^{2}
\psi(\vx)
\end{equation}
includes the vector potential (\vA) of the uniform magnetic field while,
\beq
V&=& \int d^{2}r  \: U(\vx) \psi^\dagger (\vx) \psi(\vx) \nonumber  \\
& & + \frac{1}{2} \int d^{2}r d^{2}r' \: \psi^{\dagger}(\vx) \psi^{\dagger}
(\vx') V(\vx - \vx') \psi(\vx') \psi(\vx)
\eeq
includes the interactions as well as any one-body potentials. The parameter
$\lambda$ serves, formally, to organize the large field expansion which is
then a small $\lambda$ expansion. The restriction to the LLL is acheived
by constructing a hermitian generator $T$ such that,
\begin{equation}
\widetilde{H}= e^{i T} H e^{-iT}  \label{eq:htwid}
\end{equation}
has no matrix elements connecting purely LLL states with to those with some
higher LL content. At small $\lambda$, it is reasonable to assume that no
level crossings occur among states separated at zeroth order by the cyclotron
gap. Hence the eigenstates of $\widetilde{H}_L=P\widetilde{H}P$, where $P$
projects onto the LLL, are the lowest lying eigenstates of the full
$\widetilde{H}$. (To avoid littering our equations with factors of $P$, in
the following we shall often denote the restriction of any operator to the
LL, obtained by sandwiching an operator between two projectors, by the
subscript $L$.)
Finally, the matrix elements of an operator $O$ in the full Hilbert space are
reproduced by those of $\widetilde{O}_L$ where,
\begin{equation}
\widetilde{O}= e^{i T} O e^{-iT} .
\end{equation}
To summarize: we trade the fluctuations to higher LLs for modified operators.
The full solution is obtained by diagonalizing the restricted problem
obtained in this fashion.

The generator $T$ is constructed as a series in $\lambda$,
\begin{equation}
T = \sum_{k=1}^{\infty} \lambda^k T^k
\end{equation}
and yields in turn, the series
\begin{equation}
\widetilde{H} = \frac{1}{\lambda} \sum_{k=0}^{\infty} \lambda^k H^k
\label{eq:hexp}
\end{equation}
and
\begin{equation}
\widetilde{O} =  \sum_{k=0}^{\infty} \lambda^k O^k .
\label{eq:oexp}
\end{equation}
Note that the leading order terms are of order $\lambda$, $\lambda^{-1}$, and
$\lambda^0$ respectively. From Eq.~(\ref{eq:htwid}) we see that the lowest
order piece of $T$ is specified by requiring that,
\begin{equation}
\langle E | V + i [T^1,H_K] | L \rangle =0    \label{eq:tdef}
\end{equation}
for all LLL states $|L\rangle$ and states with higher LL content $|E\rangle$;
it is fully specified by the further choice $T^1_L \equiv 0$.
It follows therefore that $\widetilde{H}_L = (\hbar \omega_c/2\lambda)
\widehat{N}_L + V_L + O(\lambda)$ where $\widehat{N}$ is the number operator
and $\widetilde{O}_L = O_L + O(\lambda)$; these are simply the naive
projection onto the LLL.

\noindent
{\bf Current Operator:} We now consider the equation of continuity within
this framework.  First, a crucial point which was already made in
\cite{ref:lri}. Whereas the expansion in Eq.~(\ref{eq:oexp})
holds for the density operator $\widetilde{\rho}(\vx)$, the current operator
derives from the kinetic energy,
\begin{equation}
{\bf j}(\vx) =  \frac{1}{\lambda} \left. \frac{ \delta H_K[{\bf A}(\vx)]}{
\delta {\bf A}(\vx)} \right|_{{\bf A}={\bf A}_b}
\end{equation}
and hence its transform has a series,
\begin{equation}
\widetilde{{\bf j}}(\vx) = \frac{1}{\lambda} \sum_{k=0}^{\infty} \lambda^k
\, {\bf j}^k(\vx)
\end{equation}
that begins with a term of order $1/\lambda$ \cite{ref:fn3}. Hence continuity,
\begin{equation}
-\partial_t \widetilde{\rho}(\vx) \equiv \frac{i}{\hbar}
[\widetilde{\rho}(\vx),
\widetilde{H}]
= \nabla \cdot \widetilde{{\bf j}}(\vx)
\end{equation}
implies that the naively projected hamiltonian and density generate {\em two}
equations:
\beq
\frac{i}{\hbar} [\widetilde{\rho}_L^0(\vx),\widetilde{H}^0_L] &=& \nabla \cdot
\widetilde{{\bf j}}^0_L(\vx) \label{eq:triv} \\
\frac{i}{\hbar} [\widetilde{\rho}_L^0(\vx),\widetilde{H}^1_L] &=& \nabla \cdot
\widetilde{{\bf j}}^1_L(\vx) . \label{eq:nontriv}
\eeq
As $\widetilde{H}^0_L \propto \widehat{N}_L$, Eq.~(\ref{eq:triv}) shows that
$\widetilde{{\bf j}}^0_L(\vx) \equiv {\bf j}^0_L(\vx)$ is purely solenoidal.
We also see that for continuity to hold, we must calculate the current to
the lowest non-trivial order of the canonical transformation. Note that while
the former is a consequence of the exact degeneracy of the LLL states, the
``staggered'' structure of the continuity equation is generic to all problems
in which we project onto a subspace of the kinetic energy, perhaps in
combination with another operator such as a periodic potential. In all such
problems we need to compute the current beyond the lowest order.

Using Eq.~(\ref{eq:tdef}) it is straightforward to write the following
expression for the $O(\lambda^0)$ piece of the LLL current operator:
\begin{equation}
\widetilde{{\bf j}}^1_L(\vx) = P {\bf j}(\vx) \frac{1}{ \hbar \omega_c
\widehat{N}/2 - H_K}
(1 - P) V P  + P V  (1 - P) \frac{1}{ \hbar \omega_c \widehat{N}/2- H_K}
{\bf j}(\vx) P
\label{eq:j1def}
\end{equation}
Despite appearances, this is a purely LLL object and its magnetic field
dependence only involves the magnetic length $l$; we
have, however, failed to find a general form in which the sum over
intermediate states has been carried out.
We may verify directly that this expression obeys Eq.~(\ref{eq:nontriv}). We
begin by noting that
\beq
\nabla \cdot \widetilde{{\bf j}}^1_L(\vx) &=& i P [\rho(\vx),H_K +V]
\frac{1}{\hbar \omega_c \widehat{N}/2 - H_K} (1 - P) V P \nonumber \\
&+& i P V (1-P) \frac{1}{\hbar \omega_c \widehat{N}/2 - H_K}
[\rho(\vx),H_K +V] P.
\eeq
Using $[\rho(\vx),H_K+V]=[\rho(\vx),H_K - \hbar \omega_c \widehat{N}/2]$ we
can rewrite the above as
\begin{equation}
\nabla \cdot \widetilde{{\bf j}}^1_L(\vx) = i P \rho(\vx) (1-P) V P - i P V
(1-P) \rho(\vx) P
\end{equation}
which is equivalent to Eq.~(\ref{eq:nontriv}).

We have now constructed a current operator $(1/\lambda)\widetilde{{\bf j}}^0_L
+ \widetilde{{\bf j}}^1_L$ that, along with the
naively projected Hamiltonian and density, obeys the equation of continuity.
Unfortunately, we still cannot use this operator in an unrestricted fashion.
The problem arises when we try to consistently compute matrix elements to
$O(\lambda^0)$ and are forced to keep track of the evolution of the states
themselves in response to the $O(\lambda)$ term in $\widetilde{H}_L$. For
example, if $|\alpha,L\rangle$ are the eigenstates of $\hbar \omega_c
\widehat{N} /2 + V_L$
with eigenvalues $\epsilon_\alpha$ and $|\tilde{\alpha},L\rangle$ the
corresponding eigenstates of $\widetilde{H}_L$ then,
\beq
\langle \tilde{\alpha} | \widetilde{{\bf j}}_L(\vx) |  \tilde{\beta} \rangle
&=& \langle \alpha | (1/\lambda)\widetilde{{\bf j}}^0_L(\vx)
+ \widetilde{{\bf j}}^1_L(\vx) | \beta \rangle  \nonumber \\
&+& \sum_{\gamma \ne \alpha} \frac{ \langle \alpha | \widetilde{H}_L^2
| \gamma \rangle \langle \gamma |\widetilde{{\bf j}}^0_L (\vx) | \beta \rangle}
{\epsilon_\alpha - \epsilon_\gamma} \nonumber \\
&+& \sum_{\gamma \ne \beta} \frac{ \langle \alpha | \widetilde{{\bf j}}^0_L
(\vx)
| \gamma \rangle \langle \gamma |\widetilde{H}_L^2 | \beta \rangle}
{\epsilon_\beta - \epsilon_\gamma}
\label{eq:fullj}
\eeq
The extra terms are, as advertised, solenoidal and of a form that require that
we diagonalize $V_L$ before we can compute them. We remark that in the full
Hilbert space these terms arise from the computation of the first order
pieces of the degenerate subspace wavefunctions---in contrast to the
non-degenerate case these involve a contribution from within the subspace
as well \cite{ref:kramers}.

\noindent
{\bf Noether Current:} We now show that the LLL Noether current differs
from our exact high field answer deduced from Eq.~(\ref{eq:fullj}). We do
this by comparing explicit forms for the soluble problem of a linear
potential
\begin{equation}
V = - E \int d^{2}r  \: y \,  \psi^\dagger(\vx) \psi(\vx)
\label{eq:linear}
\end{equation}
The wavefunctions for this problem, in Landau gauge ${\bf A}=-B y \hat{x}$ and
after scaling lengths by $l$, times by $\omega_c^{-1}$ and energies by
$\hbar \omega_c$, are
\begin{equation}
\psi^E_{ok} (\vx) = e^{i k x} \frac{e^{-(y+k-\lambda E)^2/2}}{\pi^{1/4}}
\end{equation}
and they carry a current
\beq
{\bf j}^E_{ok} (\vx) &=& \hat{\bf x} \, \frac{1}{\lambda} (y+k)
{e^{-(y+k-\lambda E)^2} \over \sqrt{\pi}} \nonumber \\
&=& \hat{\bf x} \, (y+k) {e^{-(y+k)^2} \over \sqrt{\pi}} \left[
\frac{1}{\lambda} + 2 E (y+k) + O(\lambda) \right]
\label{eq:lincurrent}
\eeq
The expansion to $O(\lambda)$ can also be derived directly from
Eq.~(\ref{eq:fullj}). In contrast the Noether current of Martinez and
Stone \cite{ref:ms,ref:rajaraman} for a potential $U(z,\overline{z})$ is,
in their complex co-ordinate notation:
\beq
j^{\overline{z}}(z) &=& -i \sum_{n=1}^{\infty} \frac{2^n}{n!}
\partial^{n-1}_{\overline{z}} \left[ \rho(z) \partial^{n}_z U(z,\overline{z})
\right] \nonumber \\
j^{z}(z) &=& +i \sum_{n=1}^{\infty} \frac{2^n}{n!}
\partial^{n-1}_{z} \left[ \rho(z) \partial^{n}_{\overline{z}} U(z,\overline{z})
\right]
\eeq
where we have ignored the $O(\lambda^{-1})$ purely projected piece. In our
problem, $U= E (\overline{z} - z)/2 i$ and hence,
\beq
{\bf j}^N(z) &=& \hat{\bf x} \, (\frac{j^{z} + j^{\overline{z}}}{2}) +
\hat{\bf y} \, (\frac{j^{z} - j^{\overline{z}}}{2 i}) \nonumber \\
&=& \hat{\bf x} \, E \rho(z) .
\eeq
Now, within the LLL and in Landau gauge, there is precisely one state with
a given x-momentum,
\begin{equation}
\psi_{ok} (\vx) = e^{i k x} {e^{-(y+k)^2/2} \over \pi^{1/4}}
\end{equation}
which is therefore also an eigenstate of Eq.~(\ref{eq:linear}). Its Noether
current is
\begin{equation}
{\bf j}^N_{ok} (\vx) = \hat{\bf x} \, E {e^{-(y+k)^2} \over \sqrt{\pi}}
\end{equation}
which is evidently different (see Fig.~1) from the $O(E)$ piece of the
exact result in Eq.~(\ref{eq:lincurrent}). However, this difference is local;
upon integration they both yield the drift current of a single particle.

It is also instructive to compare the various expressions for the currents
when the states with $-\infty \le k < 0$ are filled---a situation that
approximates the edge of an ideal $\nu=1$ state in a more realistic
geometry. These have the forms ($\rm erfc$ is the complementary error
function \cite{ref:as}),
\beq
\widetilde{{\bf j}}_L^1(\vx) &=& \hat{\bf x} \, \frac{E}{4 \pi} \left[
{\rm erfc}(-y) - \frac{2}{\sqrt{\pi}} y e^{-y^2} \right] \nonumber \\
{\bf j}^N(\vx) &=& \hat{\bf x} \, \frac{E}{4 \pi} {\rm erfc}(-y)
\eeq
for the $O(\lambda^0)$ currents, and
\begin{equation}
{\bf j}^E(\vx) = \hat{\bf x}\, \left[ \frac{E}{4 \pi} {\rm erfc}(-E-y) +
\frac{e^{-(y-E)^2}}{4 \pi^{3/2} } \right]
\end{equation}
for the full current inclusive of the $O(\lambda^{-1})$ piece, and are
shown in Fig.~2. These illustrate the general features that the drift of a
uniform fluid in a slowly varying potential is represented correctly by
both $O(\lambda^0)$  forms but that there
is additional structure in the exact answer in regions of non-uniform
density.
An amusing feature of Fig.~2 is that the $O(\lambda^{-1})$
piece of the exact answer dominates the drift, and even changes the sign of
the edge current; however, this contribution is non-dynamical and is {\em
not} the subject of edge state theory \cite{ref:wen}.

\noindent
{\bf Discussion:} We begin by stating once more the problem that we have
attempted to solve. In the high field limit it is clear that
the eigenstates are constructed dominantly out of states in the LLL.
However, as the states in the LLL are degenerate in kinetic energy the
matrix elements of $\partial_t \rho (\vx) \equiv (i/\hbar)
[H_K + V, \rho(\vx)]$ between them vanish,
and hence cannot describe non-solenoidal current flows. The actual
evaluation of the current shows that even among solenoidal flows \cite{ref:fn6}
we get only the trivial piece arising from density gradients \cite{ref:gmp}.
The issue then is one of obtaining a formal description of the high field
limit that allows for movement of charge, and in particular for the drift
currents that flow in response to applied potentials, e.g. in a measurement
of the Hall conductance!

The most straightforward procedure is to explicitly keep track of the pieces of
the wavefunctions in higher LLs that are evidently necessary for this task. A
more elegant solution is to note \cite{ref:gmp} that if we restrict even the
intermediate states in the evaluation of $[H_K +V, \rho(\vx)]$ to the LLL, its
matrix elements {\em no longer} vanish. Finally, we rewrite the commutator
as $\nabla \cdot {\bf j}^N(\vx)$ \cite{ref:rajaraman} to obtain a purely LLL
current operator. The current operator obtained in this fashion or equivalently
by the Noether construction of Martinez and Stone \cite{ref:ms} is, however,
ambiguous upto solenoidal terms.

Our canonical transformation analysis was intended to clarify two issues.
First, by explicitly keeping track of the entire dynamics in a high-field
expansion we hope we were able to make the mildly magical alteration of
the commutator somewhat more palatable. Second, as there is a current
operator in the full theory whose form is unambiguous on physical grounds,
we wanted to check if the Noether current is its high field limit and we
have concluded that it is not. Though the difference is solenoidal it
produces structure on the scale of $l$ and hence {\em cannot} arise naturally
in a purely LLL formalism.
As we have already remarked, for nearly uniform flows in slowly varying,
possibly internally generated, potentials both Eq.~(\ref{eq:lincurrent})
and the Noether current reduce to the
semi-classical drift---in other situations, however, the former is the
correct answer \cite{ref:caveat}.

%ACKNOWLEDGEMENTS
\acknowledgements
We thank S. Kivelson, M. Gelfand, A. Leggett, J. Martinez and M. Stone for
useful discussions. One of us (RR) would like to acknowledge the hospitality
of the University of Illinois at Urbana-Champaign. This work was supported
in part by NSF grants, nos. DMR 91--22385 and DMR 91--57018.

% REFERENCES

%FIGURE CAPTIONS

\begin{figure}
\caption{
Current profiles in eigenstates of the linear potential for $E=0.1$: high
field limit (solid) and LLL Noether construction (dashed), both without the
purely projected contribution. The corresponding probability density, in
units of $1/l^2$, is numerically simply $10$ times the Noether current.
}
\label{fig1}
\end{figure}

\begin{figure}
\caption{
Current profiles at the non-interacting $\nu=1$ edge for $E=0.1$: high
field limit (solid), LLL Noether construction (dashed), both without the
purely projected contribution, and the exact answer (short-dashes),
including the latter. The density is $10$ times the Noether current.
}
\label{fig2}
\end{figure}


\begin{references}
\bibitem{ref:fn6} We note that any time independent flow is solenoidal.
\bibitem{ref:caveat} Of course, for calculations which are insensitive to
the solenoidal ambiguity, such as the derivation of the edge algebra in
\cite{ref:ms}, the Noether form can be useful.
\end{references}
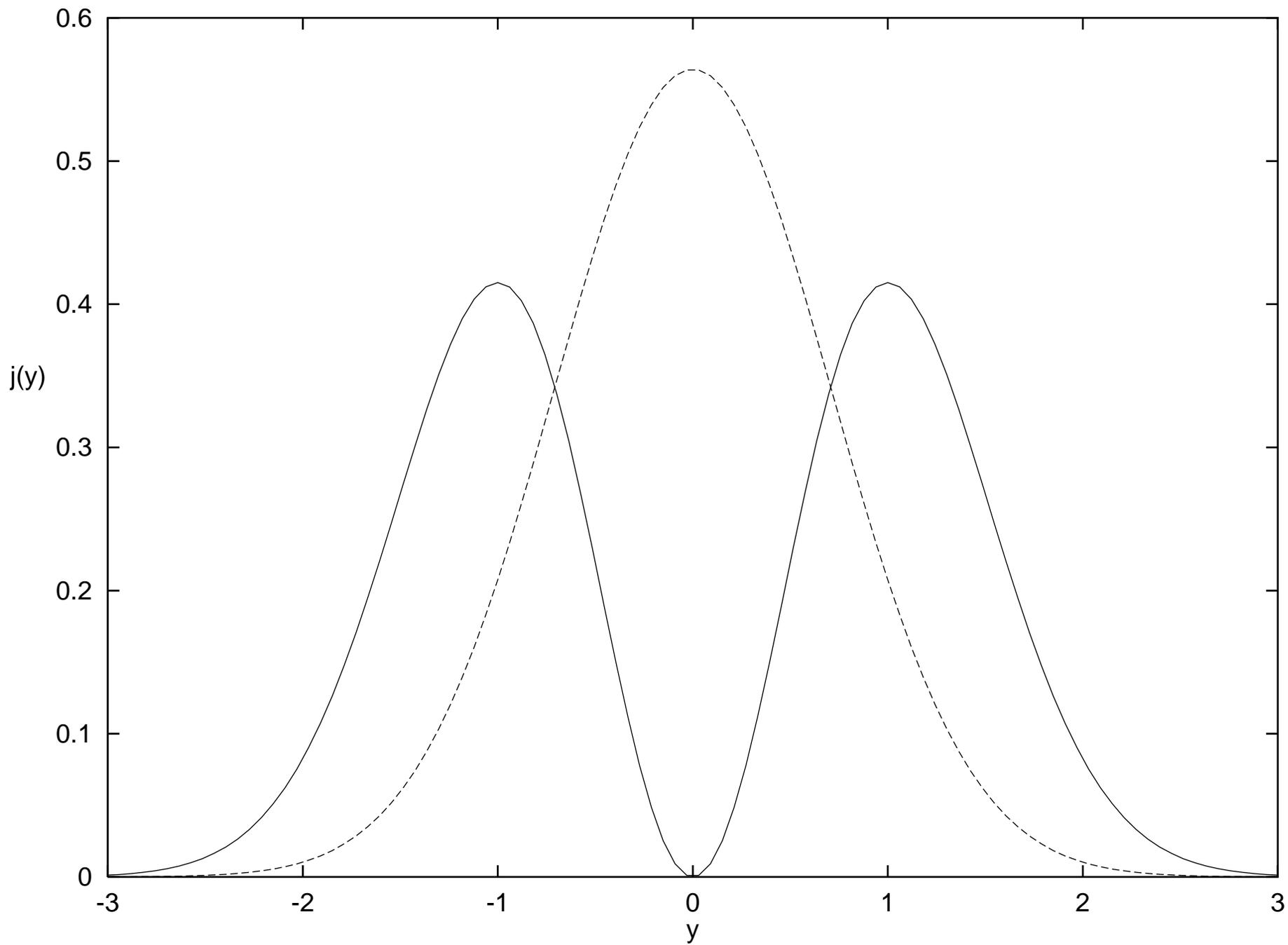
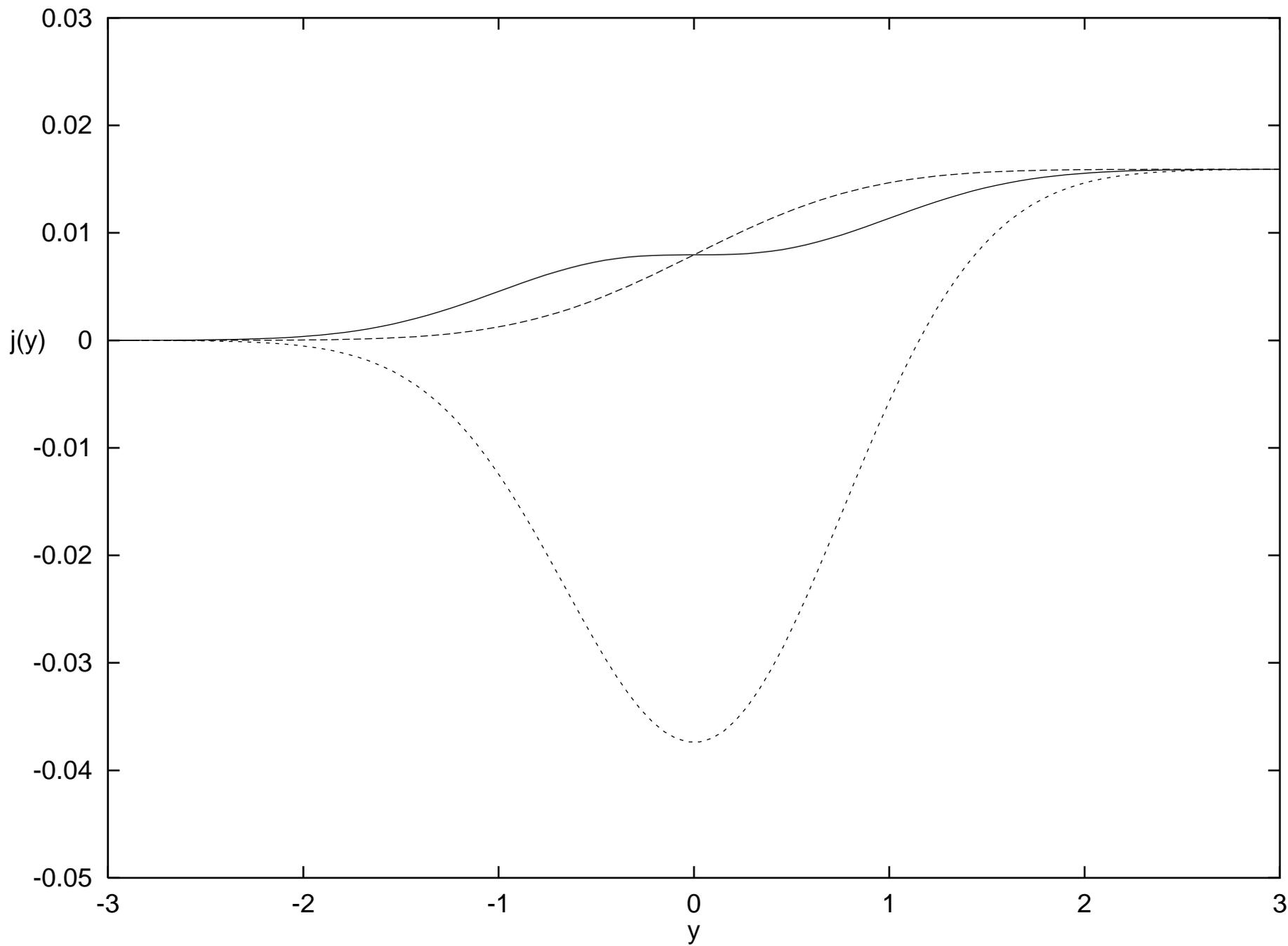
\end{document}